# Bayesian Odds-Ratio Filters:
# A Template-Based Method for Online Detection of P300 Evoked Responses

Asim M. Mubeen[a], Kevin H. Knuth[a,b,c]

[a]*Knuth Cyberphysics Lab, Department of Physics, University at Albany, Albany NY, USA*
[b]*Department of Informatics, University at Albany, Albany NY, USA*
[c]*Autonomous Exploration Inc., Andover MA , USA*

**Abstract:** Template-based signal detection most often relies on computing a correlation, or a dot product, between an incoming data stream and a signal template. While such a correlation results in an ongoing estimate of the magnitude of the signal in the data stream, it does not directly indicate the presence or absence of a signal. Instead, the problem of signal detection is one of model-selection. Here we explore the use of the Bayesian odds-ratio (OR), which is the ratio of posterior probabilities of a signal-plus-noise model over a noise-only model. We demonstrate this method by applying it to simulated electroencephalographic (EEG) signals based on the P300 response, which is widely used in both Brain Computer Interface (BCI) and Brain Machine Interface (BMI) systems. The efficacy of this algorithm is demonstrated by comparing the receiver operating characteristic (ROC) curves of the OR-based (logOR) filter to the usual correlation method where we find a significant improvement in P300 detection. The logOR filter promises to improve the accuracy and speed of the detection of evoked brain responses in BCI/BMI applications as well the detection of template signals in general.

*Index Terms*— biomedical signal processing, digital signal processing, medical signal detection, Bayesian, Bayesian model testing

## I. INTRODUCTION

Neurophysiology in general, and brain machine interface (BMI) in particular, relies on the detection and characterization of the electric and magnetic field potentials produced by the brain in response to sensory stimulation or in association with its cognitive and/or motor operations and planning. These potentials originate from the transmembrane current flow produced by multiple ensembles of hundreds of thousands of synchronously firing neurons [1], [2]. Human scalp electroencephalographic (EEG) recording has the advantage of being noninvasive, inexpensive and portable, which make it a very popular technique among the BCI/BMI community.

Event-related potentials (ERPs) are evoked brain responses synchronized to sensory, cognitive and motor events [3]. As such, they consist of relatively reproducible waveshapes embedded in background EEG activity [4], [5]. In particular, the P300 evoked potential [6], [7] is a positive peak that is evoked 300 ms after stimulus onset. The P300 is widely used in both BCI and BMI applications, but recent studies show that it is also useful in the diagnosis of neurological disorders and lie detection. Despite its wide use, detection of the P300 component in single-trial recordings remains challenging. More often, multiple trials are needed to detect the P300 component in ongoing EEG activity, which decreases the overall speed of BCI/BMI systems.

In general, the detection of an ERP is made difficult by its low signal-to-noise (SNR) ratio compared to the ongoing background EEG. The most commonly-used method to estimate the ERP is coherent averaging, which averages a large number of time-locked epochs of the identical stimuli presented to the subject. This technique is used in online applications of BCI [8], [9]; however, it suffers from several drawbacks [10]. One drawback is that coherent averaging implicitly assumes that the ERP waveshape is identical from trial-to-trial. This is known to be a poor assumption [11], [12] and, in fact, such trial-to-trial variability is utilized in some advanced source separation methods [4], [13], [14].

Several detection methods work by correlating a template signal with the ongoing EEG. For example, the Woody filter [15], [16], performs a correlation by taking the dot product of the ongoing EEG signal with the signal template. Other correlation-based classification techniques like Pearson's correlation method (PCM), Fisher's linear discriminant (FLD), stepwise linear discriminant analysis (SWLDA), linear support vector machine (LSVM), and the Gaussian kernel support vector machine (GSVM) have been presented and compared [17], [18]. These methods, which convert the problem of detecting P300 into a binary search problem (P300 present or not-present), have been widely used in BCI applications. In this paper we present a Bayesian odds ratio-based (OR-based) technique, which relies on a signal template to detect the P300 response in ongoing EEG. The OR-based detection method is not specific to EEG signals and can be applied to any kind of template-based signal detection. The technique is demonstrated by applying it to synthetic P300 responses. We compare our results to the template correlation (dot product / Woody filter [16]) method and demonstrate efficacy by comparing the resulting receiver operating characteristic (ROC) curves [19], [20], [21].

## II. THE LOG ODDS-RATIO FILTER

*A. Bayes' Theorem*

Bayes' theorem transforms the problem of signal detection into one of model selection where the probabilities of the considered models can be computed and compared. The posterior probability $P(m|D,M,I)$ gives us the probability of the parameter values, *m*, of the model *M*, given the recorded data *D* and relevant prior information *I*

$$P(m|D,M,I) = P(m|M,I) \frac{P(D|m,M,I)}{P(D|M,I)}. \quad (1)$$

The posterior probability $P(m|D,M,I)$ depends on both the prior probability of the model parameter values $P(m|M,I)$ and the data-dependent ratio of the likelihood of the data given the model and its specific model parameter values $P(D|m,M,I)$ to the evidence $P(D|M,I)$, which represents the probability that the data could have resulted from the model irrespective of the specific model parameter values. Since summing the posterior probability over all possible model parameter values results in unity, we can write the evidence as

$$P(D|M,I) = \int dm\ P(m|M,I)P(D|m,M,I), \quad (2)$$

which demonstrates the reason that this quantity is also referred to as the marginal likelihood.

We consider the problem of signal detection as a model selection problem where we compare the evidence provided by the data given one model to the evidence of the data provided by the data given another model. As we are trying to detect the P300 signal from ongoing EEG activity, we will refer to the model $M_N$ as noise-only (background EEG activity)

$$M_N: \quad x_m(t) = \eta_m(t), \quad (3)$$

where $x_m(t)$ is the time series data recorded from the $m^{\text{th}}$ electrode and $\eta_m(t)$ is the time series of ongoing EEG activity recorded at each channel. We refer to such activity as noise and assume it to be independent in each channel.

Similarly, we refer to the model $M_{S+N}$ as signal (P300) plus noise (background EEG activity) and write the model as

$$M_{S+N}: \quad x_m(t) = C_m \alpha s(t) + \eta_m(t), \quad (4)$$

where $x_m(t)$ is the time series data recorded from the $m^{th}$ electrode, $C_m$ is the coefficient that couples the signal source to the $m^{th}$ electrode, $\alpha$ is the amplitude of the signal in the given trial. We assume that the coefficients $C_m$ and the source signal waveshape $s(t)$ are both known. The value of the single-trial amplitude parameter $\alpha$ in the signal-plus-noise model is assumed to be between 0 and some maximum value $a$, so that $0 \leq \alpha \leq a$.

*B. Odds Ratio*

To compare the two models, we compute the odds ratio (*OR*), which is the ratio of the marginal likelihoods, or evidences, of each model. However, in this problem, the noise-only model $M_N$ has no model parameters. So we compare the marginal likelihood of $M_{S+N}$ to the likelihood $M_N$:

$$OR = \frac{P(D|M_{S+N},I)}{P(D|M_N,I)} \equiv \frac{Z_{S+N}}{Z_N} \quad (5)$$

where

$$Z_N = P(D|M_N,I) = P(x(t)|C,\eta(t),I) \quad (6)$$

And

$$Z_{S+N} = P(D|M_{S+N},I) = \int_0^a d\alpha\, P(\alpha|I)\, P(x(t)|C,\eta(t),\alpha,I) \quad (7)$$

We assign a Gaussian likelihood to both models, and note that this does not necessarily mean that the noise is Gaussian, but rather that the expected squared deviation $\sigma_\eta^2$ from the mean is a relevant quantity. We then have that (6) becomes

$$Z_N = (2\pi\sigma_\eta^2)^{-MT/2} \exp\left[-\frac{1}{2\sigma_\eta^2}\left(\sum_{m=1}^{M}\sum_{t=1}^{T}(x_m(t))^2\right)\right] \quad (8)$$

and that the likelihood term in (7) becomes

$$P(x(t)|C,\eta(t),\alpha,I) = (2\pi\sigma_s^2)^{-MT/2} \exp\left[-\frac{1}{2\sigma_s^2}\left(\sum_{m=1}^{M}\sum_{t=1}^{T}(x_m(t) - C_m \alpha s(t) + \eta_m(t))^2\right)\right] \quad (9)$$

We also assume that the distribution of single-trial amplitudes $\alpha$ is Gaussian with mean $\hat{\alpha}$ and variance $\sigma_\alpha^2$ so that

$$P(\alpha|I) = (2\pi\sigma_\alpha^2)^{-1/2} \exp\left(-\frac{1}{2\sigma_\alpha^2}(\alpha - \hat{\alpha})^2\right) \quad (10)$$

Applying the probability assignments in (9) and (10) to the integral in (7), we find that

$$Z_{S+N} = \frac{(2\pi\sigma_s^2)^{-MT/2}}{(2\pi\sigma_\alpha^2)^{1/2}} \exp\left[\left(\frac{1}{2\sigma_s^2}\right)\left(\frac{E^2}{D} - F\right)\right] \times \left[\frac{\sqrt{\pi}\sigma_s^2}{D}\right]\left[erf\left(\frac{2\sigma_s^2(aD-E)}{D^2}\right) - erf\left(\frac{2E\sigma_s^2}{D^2}\right)\right] \quad (11)$$

where the function *erf(x)* is the *error function* defined as the integral of a Gaussian with zero mean and unit variance from zero to *x* [22].

Also by defining

$$S_\alpha^2 = \frac{\sigma_s^2}{\sigma_\alpha^2} \qquad (12)$$

we have that

$$D = \left( S_\alpha^2 + \sum_{t=1}^{T}\sum_{m=1}^{M} C_m^2 \, s^2(t) \right), \qquad (13)$$

$$E = \left( S_\alpha^2 \hat{\alpha} + \sum_{t=1}^{T}\sum_{m=1}^{M} C_m x_m(t) \, s(t) \right) \qquad (14)$$

and

$$F = \left( S_\alpha^2 \hat{\alpha}^2 + \sum_{t=1}^{T}\sum_{m=1}^{M} (x_m(t))^2 \right). \qquad (15)$$

It is often easier to work with the logarithm of the probabilities. Taking the logarithm of the odds ratio (5) we have that

$$\log OR = \log\left(\frac{Z_{S+N}}{Z_N}\right) = \log(Z_{S+N}) - \log(Z_N), \qquad (16)$$

which after some algebraic simplification, including the assumption that $\sigma_s \approx \sigma_\eta$, results in

$$\log OR = \frac{1}{2}\left(\frac{E^2}{D\sigma_s^2} - \frac{\hat{\alpha}^2}{\sigma_\alpha^2}\right) - \frac{1}{2}\log(2\pi\sigma_\alpha^2) + \log\left[\frac{\sqrt{\pi}\sigma_s^2}{D}\right] + \log\left[erf\left(\frac{2\sigma_s^2(aD - E)}{D^2}\right) - erf\left(\frac{2E\sigma_s^2}{D^2}\right)\right] \qquad (17)$$

Since the log odds ratio compares the evidence of the two models, it can be used as an index for signal detection. We will refer to this filter as the logOR filter.

The remainder of this paper focuses on evaluating the performance of this proposed method by comparing and contrasting its receiver operator characteristic (ROC) curves with those produced by the usual correlation method (dot product). These ROC curves are constructed by applying these techniques to times-series data consisting of synthetic EEG background signals in which are embedded P300 targets. The next section discusses the construction of the synthetic EEG data.

## III. SYNTHETIC EEG DATA

To analyze the performance of the logOR filter, we generated synthetic EEG data representing both EEG background and the P300 evoked response. Three channels of synthetic EEG data (Figure 1a) were generated to simulate recordings from Cz, Pz and Fz, which are commonly used for P300-based BMI applications [18]. The dipole model is used to scale the data among different electrodes. The data from each of these channels consisted of 300 epochs each being 800 ms in length and comprised of 200 samples, which is consistent with a sampling rate of 250 Hz. Thirty epochs were selected to each host a single P300 response.

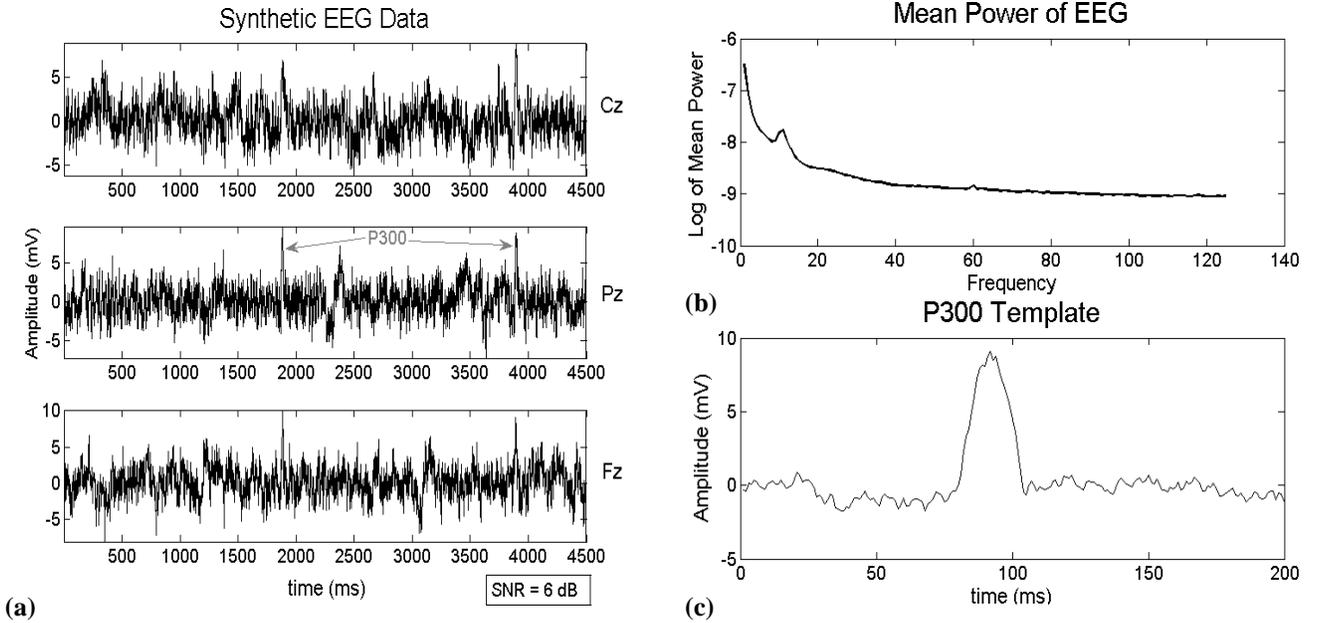

**FIGURE 1.** (a) An illustration of the synthetic EEG data generated for the filter evaluation. The three panels represent data recorded from different channels: Cz, Pz, Fz (top to bottom). Two P300s, indicated by arrows are embedded in these traces. (b) This figure illustrates the mean power spectrum of the synthetic EEG data. (c) The target signal (P300) template is shown in this figure. Note that we are using a noisy template to simulate a template generated by taking averages of P300 recording

We randomly generated a sequence of stimulus event times and used these times as latencies at which to superimpose the P300 template onto the synthetic ongoing background EEG signal, which we refer to as noise [23]. The stimulus event times were stored and used to identify as true/false-positives and true/false-negatives during the filter evaluation. The remaining 270 epochs exhibited only ongoing background EEG (noise).

The synthetic P300 waveform was represented using the classical theory [23], [24] where an ERP waveform reflects a phasic-burst of activity. The P300 template (Figure 1b) was created with by producing a peak of width of 100ms starting around 300 ms after onset. To produce a P300 template more similar to what one would generate in the lab by averaging recorded P300 responses, we added low-level background noise to the P300 template.

To generate the ongoing EEG background, we used MATLAB code provided online by Yeung and Bogacz [25]. Fifty different sine waves were superimposed to create the background signal. This was done by randomly selecting a frequency for each sine wave between 0 and 50 Hz. The phase was selected at random, and the amplitude was chosen to be consistent with the Human EEG power law spectrum on which was superimposed two peaks at 10 and 60 Hz [26] (Figure 1c).

To study the effect of the SNR on the filter performance, we created 17 data sets where the SNR, calculated by the formula

$$SNR_{dB} = 10\log_{10}\left(\frac{A_{Signal}}{A_{Noise}}\right)^2, \qquad (19)$$

was varied in integral steps from -6 dB to 7 dB as well as 10, 15 and 20 dB. This covers the typical SNR range seen in BCI EEG applications, which is approximately from -6 dB to 6 dB [27].

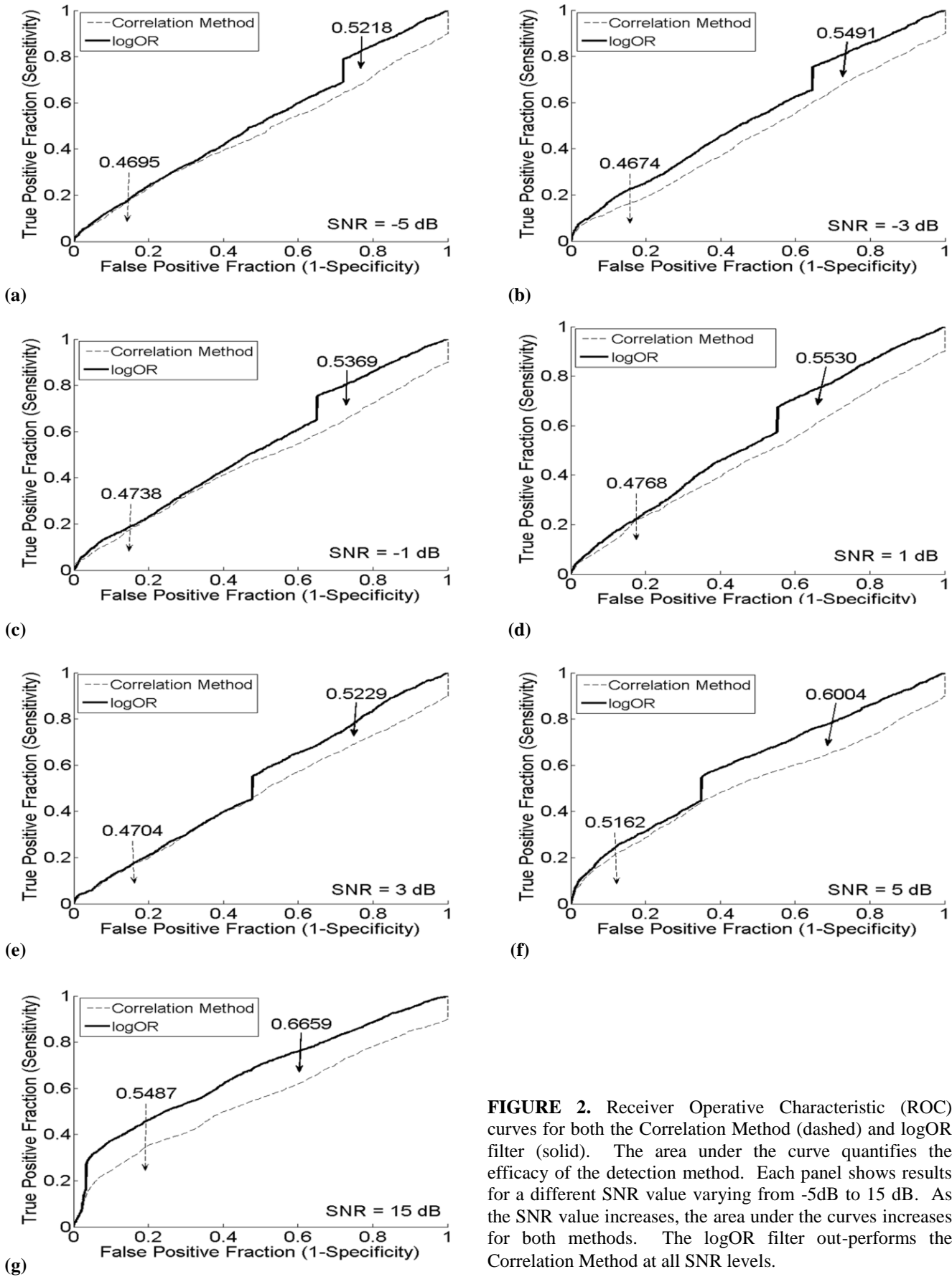

**FIGURE 2.** Receiver Operative Characteristic (ROC) curves for both the Correlation Method (dashed) and logOR filter (solid). The area under the curve quantifies the efficacy of the detection method. Each panel shows results for a different SNR value varying from -5dB to 15 dB. As the SNR value increases, the area under the curves increases for both methods. The logOR filter out-performs the Correlation Method at all SNR levels.

## IV. RECEIVER OPERATIVE CHARACTERISTICS (ROC) CURVES

ROC curves serve as a performance measure of signal detection [20], [21] by plotting the true positive rate (also known as *sensitivity*) versus the false positive rate (*1–specificity*). Sensitivity is computed by identifying the fraction of correctly detected targets (P300) and specificity refers to the fraction of non-targets (non-P300) identified as non-targets. The fractional area under the ROC curve serves to quantify the performance of the detection methods so that a perfect detection method will result in unity; whereas a completely failed method will result in zero. The result is such that the more accurate the detection method, the greater the fractional area under the ROC curve.

## V. DETECTION THRESHOLD

The selection of a detection threshold value is a difficult task. As the detection threshold increases, the sensitivity decreases while the specificity increases, which means that the false positive fraction (1–specificity) decreases. To produce ROC curves, we calculate sensitivity and (1-specificity) for each distinct value of the detection measure (i.e logOR / Correlation) to consider it as a candidate for detection cutoff. By plotting (1–specificity) versus sensitivity, the efficacy of the detection method can be quantified by the area under the ROC curve. The ROC curve starts from the most strict decision cutoff point where sensitivity and (1–specificity) values are zero where there will be no positive detection, and ends at most lenient cutoff point where sensitivity and (1–specificity) values are one and all detection values are taken to represent a positive detection. The value of the area under the curve for the best detection method should be one [28], [29]. However, when a method results in an ROC curve where the area under the curve is less than one; there is always a tradeoff between sensitivity and (1–specificity). So, the detection threshold can be found by minimizing the distance between sensitivity and specificity. One can also find the threshold value by plotting the sensitivity and specificity versus the detection criteria, and selecting the decision threshold value based on the intersection point of sensitivity and specificity curves.

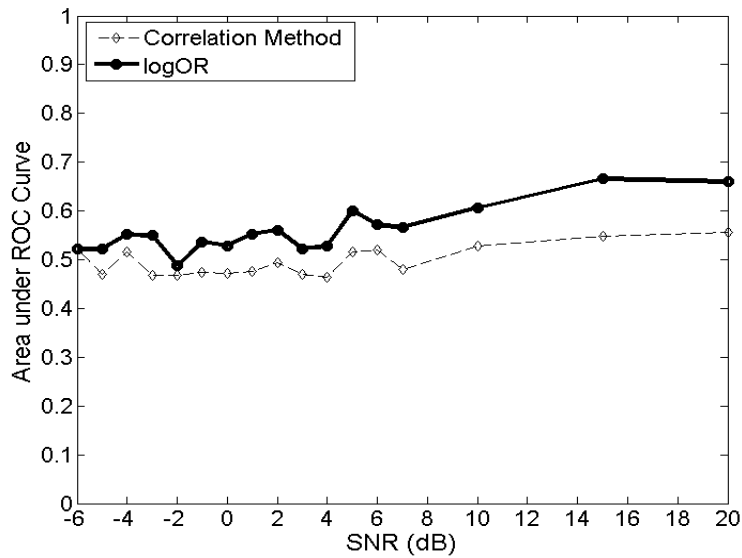

**FIGURE 3**. SNR vs. Area under ROC curves for the Correlation Method and the logOR filter. The value of Area under the curve increases as SNR increases for both methods. For higher values of SNR, the logOR filter performs much better than the Correlation Method.

## VI. RESULTS

In this paper we developed and evaluated a template-based detection method that relies on the Bayesian log odds ratio (logOR filter) to detect the presence or absence of a P300 signal in a synthetic ongoing EEG dataset representing three commonly-used EEG channels: Cz, Pz and Fz.

We applied the logOR filter to the synthetic EEG data and compared the results to those obtained using the traditional correlation filter. Using the detections and the sequence of events used to generate the data, we computed the ROC curves for different SNR ranging from -6 dB to 20 dB to evaluate the performance of both methods. Figure 2 shows the resulting ROC curves for several SNR values ranging from -5 dB to 15 dB. The area under the ROC curve is consistently greater for the logOR filter indicating that the logOR filter out-performs the traditional correlation filter. A plot of the area under the ROC curves as a function of P300 SNR can be seen in Figure 3. Both methods exhibit similar performance at the lowest SNR value, but as the SNR increases, one can see that the logOR filter out-performs the correlation-based filter.

## VII. DISCUSSION

Most template-based methods designed to detect a target signal in a continuous EEG stream rely on the cross correlation of the incoming data with the template or target signal [15], [16]. Here we present a new approach in which we recast the problem of signal detection in terms of evidence-based model selection. By computing the Bayesian log odds-ratio between two models (the signal plus noise model, $M_{S+N}$, and the noise only model, $M_N$), we produce an index, which we refer to as the logOR filter. The value of the logOR filter index is high when the target signal, for example a P300, is present and low otherwise. We studied the performance of these two methods by applying them to synthetic EEG signals with P300 targets exhibiting SNRs ranging from -6 to 20 dB. Performance was quantified by constructing the ROC curves for each method and computing the area under the ROC curve. We found that the logOR filter out-performs the correlation method. This suggests that systems aiming to minimize the number of target trials, such as P300-based BCI/BMI systems, may benefit from employing the logOR filter. However it is observed that logOR filter performance is very sensitive to differences between the assigned and true standard deviation values of the background noise and the trial-to-trial amplitude variability of the P300 response. Therefore in a practical application one must choose the values of these parameters very carefully to calculate the logOR filter value.

## ACKNOWLEDGMENTS

*The authors would like to thank Dr. Dennis J. McFarland for valuable discussions, access to EEG data and instruction regarding the BCI2000 system*